\documentclass[preprint,showpacs,preprintnumbers,amsmath,amssymb]{revtex4}
\usepackage{graphicx}
\usepackage{dcolumn}
\usepackage{bm}

\begin{document}

\preprint{}

\title{
Free initial wave packets and the long-time behavior of \\
the survival and nonescape probabilities
}

\author{Manabu Miyamoto}%
 \email{miyamo@hep.phys.waseda.ac.jp }
\affiliation{%
Department of Physics, Waseda University, 3-4-1 Okubo, Shinjuku-ku, 
Tokyo 169-8555,  Japan 
}%

\date{\today}

\begin{abstract} 
The behavior of both the survival $S(t)$ and nonescape $P(t)$ probabilities 
at long times for the one-dimensional free particle system is shown 
to be closely connected to that of the initial wave packet at small momentum. 
We prove that both $S(t)$ and $P(t)$ asymptotically 
exhibit the same power-law decrease at long times, 
when the initial wave packet in momentum representation 
behaves as $k^m$ with $m =0$ or $1$ 
at small momentum. 
On the other hand, if the integer $m$ becomes greater than $1$, 
$S(t)$ and $P(t)$ decrease in different power-laws at long times.
%
%
%
\end{abstract}

\maketitle



The study of the space-time evolution of the wave packets 
is very significant for understanding of the scattering phenomena 
and attracting many researchers in the various fields. 
It is then helpful to know 
the complete information about the free particle system. 
For the one dimensional case, 
if the Gaussian wave packet is chosen as the initial one, 
the wave packet $\psi (x,t)$ decreases asymptotically as $t^{-1/2}$ at long times. 
However, it has been recently found that 
the maximum of wave packet does not necessarily behave as $t^{-1/2}$ 
for non-Gaussian initial-wave-packet. 
In fact, a slower decrease than $t^{-1/2}$ can be found  
for the power-law tail wave packet 
\cite{Unnikrishnan,Lillo,Mendes}, and a faster decrease than  $t^{-1/2}$ can occur for 
the wave packet which vanishes at zero momentum 
\cite{Damborenea}. 
These facts remind us of a naive question of 
how the characteristics of the initial wave packet affect 
the long time behavior of the wave packet or related quantities 
such as the survival $S(t)$ and nonescape $P(t)$ probabilities. 
The former stands for the probability 
of a state still being in the initial state at a later time $t$. 
It is widely used 
for the decaying systems 
(see, e.g., Refs.\ \cite{Fonda,Nakazato,Garcia,Dijk} and references therein). 
The latter is the probability 
to find a particle in a specific region under consideration 
at a later time $t$. 
It is also used to the decaying systems 
(see, e.g., Refs.\ \cite{Garcia,Nussenzveig,Muga,Dijk}). 
Such a question for $S(t)$ 
was answered in the following sense. 
For a one dimensional free particle system, 
it was shown that $S(t)$ behaves asymptotically like $t^{-2m-1}$, 
when the initial wave packet in momentum representation 
behaves like $k^m$ near the zero momentum $k=0$ 
with an arbitrary nonnegative integer $m$ 
\cite{Miyamoto}. 
Hence, as is pointed out in Ref.\ \cite{Damborenea}, 
the small-momentum behavior of initial wave packet 
has a crucial role to determine the long-time behavior of the survival probability. 
However, such strict structures for the wave packet and the nonescape probability 
have not been clarified completely.

In this work, we consider the asymptotic behavior 
of a free wave packet $\psi (x,t)$ at long times for the one dimensional case,  
assuming as in Ref.\ \cite{Miyamoto} that the initial wave packet 
behaves like $k^m $ at small momentum.  
The asymptotic behavior is evaluated at every position $x$ 
unlike the studies in Refs.\ \cite{Unnikrishnan,Lillo,Mendes,Damborenea}. 
This advantage enables us to discuss whether the asymptotic behavior of 
the wave packet has the position dependence. 
In addition, we are able to calculate explicitly 
the asymptotic form of not only $S(t)$ but also $P(t)$ at long times \cite{SP}. 
We then examine and clarify 
the difference between the long time behavior of the $S(t)$ and $P(t)$, 
according to the small-momentum behavior of the initial states. 
Remark that a comparison between the long time behavior of $S(t)$ and $P(t)$ was 
already made in Ref.\ \cite{Garcia} for the potential systems in another context, 
though the analysis therein was correct for $S(t)$, but not for $P(t)$. 
The correct result for $P(t)$ turned out to be the $t^{-3}$ behavior 
(see, Ref.\ \cite{controversy} and references therein).


For the one-dimensional free particle system 
with the Hamiltonian $H_0 = 
-(\hbar^2 /2M) d^2 /dx^2 $, 
we here define the survival probability $S(t)$ of 
the initial state (wave packet) $\psi$ as 
\begin{equation}
S(t):= | \langle \psi , e^{-it H_0 /\hbar } \psi \rangle |^2 
= 
\left| 
\int_{-\infty}^{\infty} \overline{\psi (x)} \psi (x,t) dx 
\right|^2 , 
\label{eqn:3}
\end{equation}
where $\psi (x,t) =  (e^{-it H_0 /\hbar } \psi )(x) $ and 
the bar ( $\bar{ }$ ) denotes complex conjugate. 
$\psi (x)$ is assumed to be square integrable. 
$S(t)$ is the probability that the state at a later time $t$ 
is found in the initial one. 
We also define the nonescape probability $P(t)$ 
as the probability that 
a particle initially prepared in the state $\psi$ 
is found in a bounded interval $[a,b]$ on the line at a later time $t$,   
\begin{equation}
P(t):= \int_a^b |\psi (x,t)|^2 dx.  
\label{eqn:6}
\end{equation}

In order to estimate the asymptotic behavior of $\psi (x,t)$, 
we first refer to the explicit solution to the Schr{\" o}dinger equation, 
\begin{eqnarray}
\psi (x,t) 
&=&  
\frac{1}{\sqrt{2\pi} } \int_{-\infty}^{\infty} e^{ikx} 
e^{-i t \hbar k^2 /2M} \hat{\psi} (k) dk, 
\label{eqn:360} \\ 
&=&  \left(\frac{M}{2\pi i\hbar t} \right)^{1/2}
\int_{-\infty}^{\infty} e^{iM|x-y|^2 / 2 \hbar t} \psi (y) dy ,   
\label{eqn:30}  
\end{eqnarray}
where $k=p/\hbar$ and 
the $\hat{\psi}(k)$ is the initial state in momentum representation. 
To see the long time behavior of the solution, 
one can consider the asymptotic expansion of the integral 
in Eq.\ (\ref{eqn:360}), using the phase stationary method \cite{Olver} 
as used in Ref.\ \cite{Lillo,Damborenea} or 
making an integration by parts for the Fourier integral \cite{Copson}. 
Then the differential coefficients of $\hat{\psi}(k)$ at $k=0$ naturally appear. 
However, to take account of the $x$-dependence in the asymptotic behavior of 
$\psi (x,t)$, 
it may be convenient for us to start with Eq.\ (\ref{eqn:30}). 
Indeed, 
expansion of the exponential function in Eq.\ (\ref{eqn:30}) immediately 
leads to the asymptotic behavior of $\psi (x,t)$ with the $x$-dependence. 
It reads 
\begin{equation}
\psi (x,t) \sim  \sum_{j=0} ^{\infty} 
\frac{(-1)^{j-1} \Gamma (j+1/2)}{\pi (i \hbar t/2M)^{j+1/2}} (G_{2j} \psi )(x) , 
\label{eqn:70}
\end{equation}
where $G_j$ is the integral operator \cite{Gj} defined by 
\begin{equation}
(G_j \psi )(x):= -\frac{1}{2(j!)} 
\int_{-\infty}^{\infty} |x-y|^j \psi (y) dy . 
\label{eqn:60} 
\end{equation}
Here we assume the exchange of the order of summation and integration to be allowed. 
Note that 
$(G_{2j} \psi )(x) $ in Eq.\ (\ref{eqn:70}) can be described  
in terms of the differential coefficient of $\hat{\psi} (k)$ at $k=0$ as same as 
the result reached from Eq.\ (\ref{eqn:360}). 
This is seen from the following formal expansion of $\hat{\psi} (k)$, 
\begin{eqnarray}
\hspace*{-5mm}
\hat{\psi}(k) 
&=& 
\frac{1}{\sqrt{2\pi} } \int_{-\infty}^{\infty} e^{-iky} \psi (y) dy 
\nonumber \\ 
&\sim&  
\sum_{j=0}^{\infty} \frac{(-ik)^{j}}{\sqrt{2\pi} j!} 
\int_{-\infty}^{\infty}
y^j \psi (y) dy 
= \sum_{j=0}^{\infty} \frac{k^{j}}{j!} \hat{\psi}^{(j)}(0) , 
\label{eqn:200}
\end{eqnarray}
where $\hat{\psi}^{(0)}(0)=\hat{\psi}(0)$. This implies that 
\begin{equation}
\hat{\psi}^{(j)}(0) = \frac{d^j \hat{\psi}(k) }{dk^j } \biggr|_{k=0}
= \frac{(-i)^{j}}{\sqrt{2\pi}} 
\int_{-\infty}^{\infty}
y^j \psi (y) dy . 
\label{eqn:210} 
\end{equation}
Then, we can rewrite the $(G_{2j} \psi )(x)$ in Eq.\ (\ref{eqn:60}) for $j=2j$ as 
\begin{equation}
(G_{2j} \psi )(x) = -\frac{\sqrt{2\pi}}{2[(2j)!]} 
\sum_{n=0}^{2j} {{2j}\choose{n}} i^{n} 
\hat{\psi}^{(n)}(0) (-x)^{2j-n}. 
\label{eqn:220} 
\end{equation}
Substituting Eq.\ (\ref{eqn:70}) 
into (\ref{eqn:3}) and (\ref{eqn:6}), 
we can obtain the asymptotic behaviors of $S(t)$ and $P(t)$ at long times as, 
\begin{equation}
S(t) \sim 
\left| 
\sum_{j=0}^{\infty} 
\frac{(-1)^{j-1} \Gamma (j+1/2)}{\pi (i \hbar t/2M)^{j+1/2}} 
\langle \psi , G_{2j} \psi \rangle 
\right|^2 ,
\label{eqn:240}
\end{equation}
and 
\begin{equation}
P(t) \sim  \int_a^b 
\left| \sum_{j=0}^{\infty} 
\frac{(-1)^{j-1} \Gamma (j+1/2)}{\pi (i \hbar t/2M)^{j+1/2}} (G_{2j} \psi )(x) 
\right|^2 dx 
\label{eqn:250}
\end{equation}
respectively. In Eq.\ (\ref{eqn:240}), 
$\langle \psi , G_{2j} \psi \rangle $ is also described in terms of 
the differential coefficients $\hat{\psi}^{(j)} (0)$ as, 
\begin{equation}
\langle \psi , G_{2j} \psi \rangle = \frac{(-1)^{j-1} \pi}{(2j)!} 
\sum_{n=0}^{2j} {{2j}\choose{n}} 
\overline{ \hat{\psi}^{(2j-n)}(0)} \hat{\psi}^{(n)}(0) . 
\label{eqn:260} 
\end{equation}

We now consider such a special case that 
the initial wave packet $\psi (x)$ satisfies 
\begin{equation}
\hat{\psi} (k) = O(k^{m}) ~~~ \mbox{as } k \rightarrow 0, 
\label{eqn:290} 
\end{equation}
where $m=1, 2, \ldots.$ 
We notice from Eq.\ (\ref{eqn:200}) that the condition (\ref{eqn:290}) 
is equivalent to the condition
\begin{equation}
\hat{\psi}^{(j)}(0) = 0, ~~~ \mbox{ for } j=0,1,\ldots, m-1. 
\label{eqn:280} 
\end{equation}
Note that the condition (\ref{eqn:280}) causes 
the $S(t)$ to behave like $t^{-2m-1}$. 
To confirm this assertion, it suffices to show that the condition (\ref{eqn:280}) 
implies the next condition
\begin{equation}
\langle \psi , G_{2j} \psi \rangle = 0, ~~~ \mbox{ for } j=0,1,\ldots, m-1, 
\label{eqn:270} 
\end{equation}
and vice versa \cite{Miyamoto}. 
In fact, substitution of Eq.\ (\ref{eqn:270}) into (\ref{eqn:240}) 
surely leads to $S(t)\sim t^{-2m-1}$. 
We briefly show the equivalence between 
the conditions (\ref{eqn:280}) and (\ref{eqn:270}). 
The fact that Eq.\ (\ref{eqn:280}) implies Eq.\ (\ref{eqn:270}) 
follows straightforwardly from Eq.\ (\ref{eqn:260}). 
Conversely, if Eq.\ (\ref{eqn:270}) holds, 
we have from $\langle \psi , G_{0} \psi \rangle = 0$ 
that $\hat{\psi}^{(0)}(0) = 0$ [see Eq.\ (\ref{eqn:260})].  
Then, we also have from 
$\hat{\psi}^{(0)}(0) = 0$ and $\langle \psi , G_{2} \psi \rangle = 0$ 
that $\hat{\psi}^{(1)}(0) = 0$ [see Eq.\ (\ref{eqn:260}) again]. 
As the same way, we can recursively show Eq.\ (\ref{eqn:280}), 
and the proof is completed. 
Under the condition (\ref{eqn:280}), we see that the first non-vanishing term 
$\langle \psi , G_{2m} \psi \rangle$ 
in the summation in Eq.\ (\ref{eqn:240}) is reduced to 
\begin{equation}
\langle \psi , G_{2m} \psi \rangle = \frac{(-1)^{m-1} \pi}{(m!)^2} 
|\hat{\psi}^{(m)}(0)|^2 . 
\label{eqn:320} 
\end{equation}
Then, we obtain the asymptotic behavior for $S(t)$ as \cite{Miyamoto} 
\begin{equation}
S(t)= \frac{\Gamma (m+1/2)^2 }{(m!)^4 ( \hbar t/2M)^{2m+1} } 
|\hat{\psi}^{(m)} (0)|^4  + O(t^{-2m-2} ) .  
\label{eqn:100}
\end{equation}
Note that this formula is also seen to be valid for $m=0$. 
%
%
%
%
%
%
%
%
%
%
%
%
%
%
%
Let us now examine how the same condition (\ref{eqn:290}) [or (\ref{eqn:280})] affects 
the asymptotic behavior of $\psi (x,t)$. 
Under the condition (\ref{eqn:280}), we see that Eq.\ (\ref{eqn:220}) reads  
\begin{equation}
(G_{2j} \psi )(x) 
=0 ~~~ \mbox{ for all } x \in \mathbb{R}, 
\label{eqn:300}
\end{equation}
where $j=0,1,\ldots, \overline{m}-1$.  
The $\overline{m}-1$ is the largest integer satisfying 
$2(\overline{m}-1) \leq m-1$, and the $\overline{m}$ turns out to be 
\begin{equation}
\overline{m} = 
\left\{ 
\begin{array}{ll}
m/2 & \mbox{for even } m \\
(m+1)/2  & \mbox{for odd } m  
\end{array}
\right. . 
\label{eqn:310}
\end{equation}
Equation (\ref{eqn:300}) consequently implies that 
the asymptotic expansion (\ref{eqn:70}) for the wave packet reads 
\begin{equation}
\psi (x,t) =  
\frac{(-1)^{\overline{m}-1} \Gamma (\overline{m}+1/2)}
{\pi (i \hbar t/2M)^{\overline{m}+1/2}} 
(G_{2\overline{m}} \psi )(x) +O(t^{-\overline{m}-3/2}),  
\label{eqn:125}
\end{equation}
as $t \rightarrow \infty$. 
One also see that this formula is valid for $m=0$. 
By using Eq.\ (\ref{eqn:280}), 
the first non-vanishing term $(G_{2\overline{m}} \psi )(x)$ 
can be reduced to a simple expression as
\begin{equation}
(G_{2\overline{m}} \psi )(x) = \frac{-\sqrt{2\pi} i^{m} }{2(m!)} 
\hat{\psi}^{(m)}(0) , 
\label{eqn:340} 
\end{equation}
for even $m$, or 
\begin{equation}
(G_{2\overline{m}} \psi )(x) = \frac{-\sqrt{2\pi} i^{m+1} }{2[(m+1)!]} 
[\hat{\psi}^{(m+1)}(0) +i(m+1)x \hat{\psi}^{(m)}(0) ] , 
\label{eqn:330} 
\end{equation}
for odd $m$. 
Substituting Eq.\ (\ref{eqn:125}) into (\ref{eqn:6}), 
we can also derive the asymptotic behavior for $P(t)$:
\begin{equation}
P(t)\hspace*{-0.5mm}=\hspace*{-0.5mm}
\frac{\Gamma (\overline{m}+1/2)^2 }{\pi^2 ( \hbar t/2M)^{2\overline{m}+1} } 
\hspace*{-1mm} \int_a^b  |(G_{2\overline{m}} \psi )(x)|^2 dx 
+ O(t^{-2\overline{m}-2} ) ,  
\label{eqn:110}
\end{equation}
as $t \rightarrow \infty$. 
The above formula is also expressed in terms of the differential coefficients 
$\hat{\psi}^{(j)} (0)$, 
by using Eq.\ (\ref{eqn:340}) [or (\ref{eqn:330})]. 
%

It is worth noting that, in the case of $m$ being odd, 
there is a possibility to find a special position, denoted by $\xi_0$, where 
$(G_{2\overline{m}} \psi )(x)$ in Eq.\ (\ref{eqn:330}) 
vanishes. 
This means that at that position the $\psi(x,t)$ follows the power-law 
in the next order. 
However, this matter may be regarded as exceptional, 
because a point in the entire line has only zero-measure. 
From Eq.\ (\ref{eqn:330}), $\xi_0$ is given by 
\begin{equation}
\xi_0 = i\hat{\psi}^{(m+1)}(0) /[(m+1) \hat{\psi}^{(m)}(0)], 
\label{eqn:370}
\end{equation}
and must be real. 
We can find such a $\xi_0$ 
for the initial wave packet, e.g., $N_m k^m e^{-{a_0}^2 (k-k_0 )^2 /2 -ix_0 k} $, 
where $a_0 >0$, $k_0$, $x_0$ $\in \mathbb{R}$, 
and $N_m$ being the normalization constant. 
In this case, the $\xi_0$ is given by $x_0+i{a_0}^2 k_0$. 
Then, it becomes real if and only if $k_0=0$, which leads to $\xi_0=x_0$, 
the center of the initial wave packet. 
%
%
Note that such a special position in Eq.\ (\ref{eqn:370}) if any 
does not have an influence on the asymptotic form of $P(t)$, 
because $P(t)$ is obtained by the integral of $|(G_{2\overline{m}} \psi )(x)|^2$.

Let us now consider and compare the long time behavior of $S(t)$ and $P(t)$. 
We see from Eq.\ (\ref{eqn:310}) that 
$m$ and $\overline{m} $ are different when $m \geq 2$, 
and this fact directly affects 
the long time behavior of $S(t)$ and $P(t)$.  
When the initial state $\psi$ satisfies 
$\hat{\psi} (k) = O(k^{m})$ with an arbitrary integer $m \geq 2$,  
$S(t)$ goes asymptotically like $t^{-2m-1}$, 
whereas $P(t)$ does like $t^{-m-1}$ for even $m$ or like $t^{-m-2}$ for odd $m$. 
For a large $m$, $S(t)$ decreases much faster than $P(t)$. 
We also see that $\overline{m} = \overline{m+1}$ for odd $m$. 
This means that, in the case of an odd integer $m$, unlike $S(t)$, 
$P(t)$ decreases in the same power law under both the conditions, 
$\hat{\psi} (k) = O(k^{m})$ and $\hat{\psi} (k) = O(k^{m+1})$. 

\begin{figure}
\rotatebox{270}{
\includegraphics[width=0.3\textwidth]{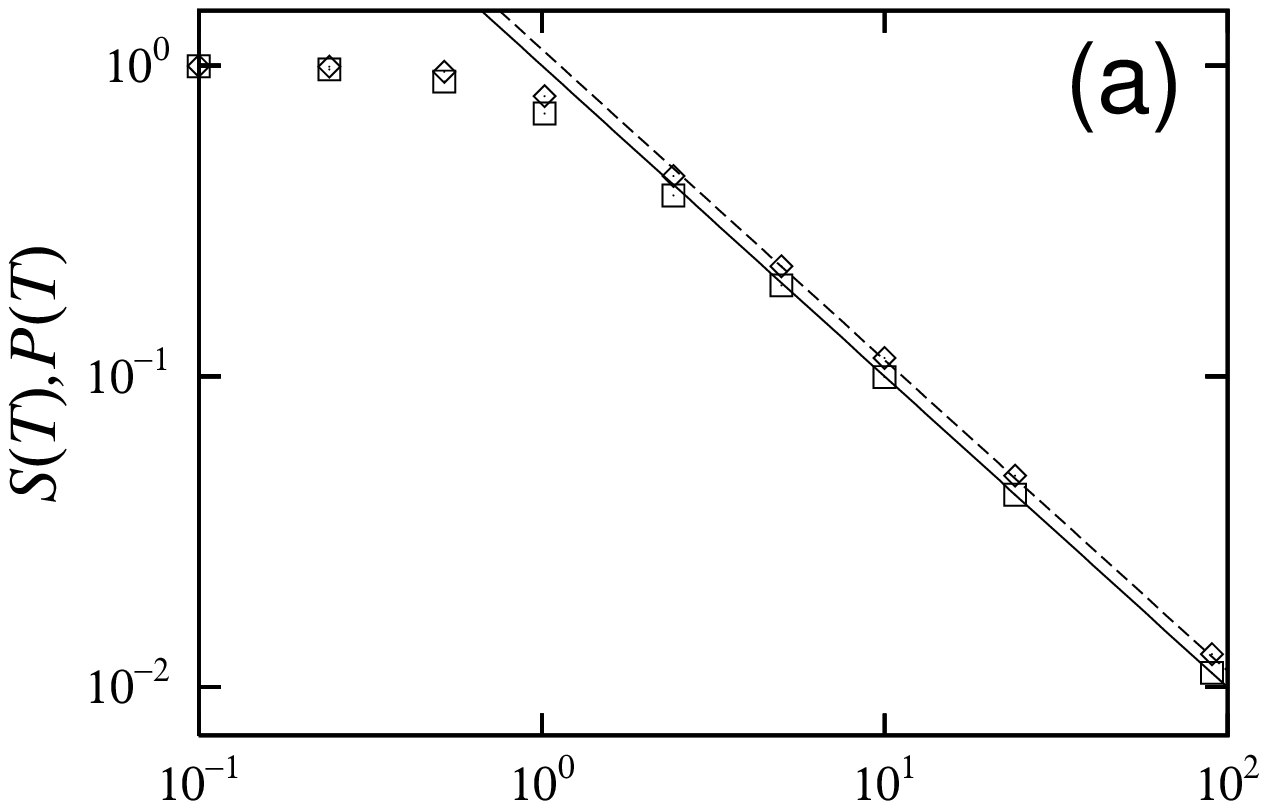}
\includegraphics[width=0.3\textwidth]{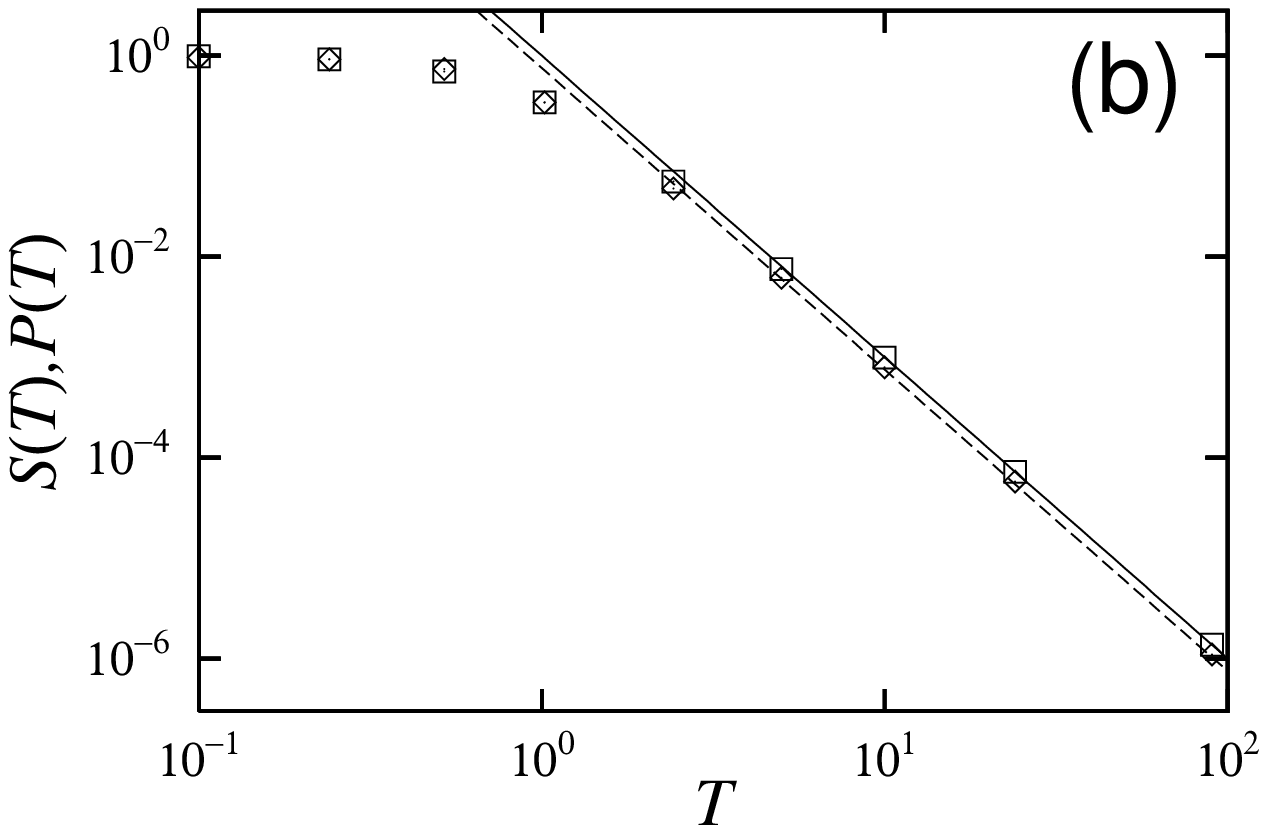}
}
\caption{\label{fig:figure1} 
(a) $S(T)$ and $P(T)$ 
(square and diamond, respectively) of the wave function  
$\phi_0$ in Eq.\ (\ref{eqn:130}), and 
their asymptotes predicted by 
Eqs.\ (\ref{eqn:100}) and (\ref{eqn:110}) 
(solid and dashed lines, respectively), 
where $T= \hbar t/2M{a_0}^2$ being the reduced time. 
In this case, $S(T)$ and $P(T)$ 
show the same power decay behavior like $T^{-1}$  at long times. 
(b) 
$S(T)$ and $P(T)$ of the wave function $\phi_1$, and 
their asymptotes. 
The notations and symbols are the same as those in (a). 
$S(T)$ and $P(T)$ exhibit the same power decay, 
however they behave like $T^{-3}$ instead of $T^{-1}$. 
Here we set $k_0 =0.0$ and $x_0 =0.0$, in $\phi_0$ and $\phi_1$,  
and $a/a_0 =-2.0$ and $b/a_0 =2.0$ in $P(T)$'s .  }
\end{figure}

\begin{figure}
\rotatebox{270}{
\includegraphics[width=0.3\textwidth]{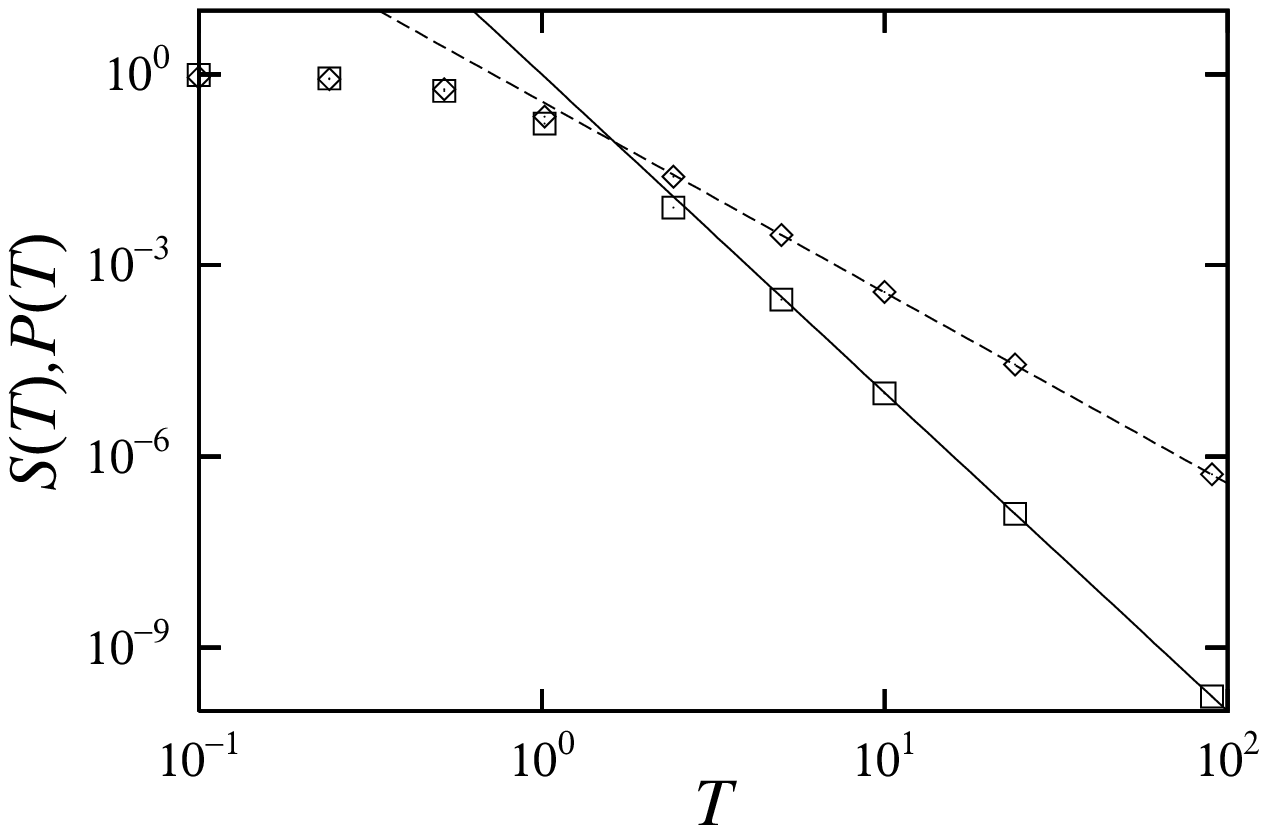}
}
\caption{\label{fig:figure2} 
$S(T)$ and $P(T)$ of the wave function $\phi_2$, and 
their asymptotes, where $T= \hbar t/2M{a_0}^2$. 
The same notations and symbols as in Fig.\ \ref{fig:figure1} are used. 
In this case, $S(T)$ and $P(T)$ 
exhibit different power decays at long times. 
The former behaves like $T^{-5}$ while the latter like $T^{-3}$. 
Here we set $k_0 =0.0$, and $x_0 =0.0$, in $\phi_2$, 
and $a/a_0 =-2.0$ and $b/a_0 =2.0$ in $P(T)$.}
\end{figure}

To illustrate the difference in the long time behavior of 
$S(t)$ and $P(t)$, 
we choose 
the three initial wave functions $\phi_0 (x)$, $\phi_1 (x)$, and $\phi_2 (x)$, 
defined by 
\begin{equation}
\hat{\phi}_m (k) = N_m k^m e^{-{a_0}^2 (k-k_0 )^2 /2 -ix_0 k}, ~~~
\mbox{ for } m=0, 1 , 2. 
\label{eqn:130}
\end{equation}
These are the same ones considered after Eq.\ (\ref{eqn:370}). 
They behave like $\hat{\phi}_m (k) = O(k^{m})$ for small $k$. 
Figure \ref{fig:figure1} 
shows the time evolution of $S(t)$ and $P(t)$, 
and their asymptotic forms predicted by 
Eqs.\ (\ref{eqn:100}) and (\ref{eqn:110}). 
The initial states $\phi_0$ and $\phi_1$ are used in 
Fig.\ \ref{fig:figure1} (a) and (b), respectively. 
It is clearly seen 
that in Fig.\ \ref{fig:figure1} (a) 
$S(t)$ and $P(t)$  
behave asymptotically like $t^{-1}$ at long times,  
and in Fig.\ \ref{fig:figure1} (b) like $t^{-3}$. 
In these cases, the difference between the behavior of 
$S(t)$ and $P(t)$ is not found. 
On the other hand, we notice 
that in Fig.\ \ref{fig:figure2} $S(t)$ and $P(t)$ 
for the initial state $\phi_2$ 
differ asymptotically at long times. 
The former behaves asymptotically like $t^{-5}$, however 
the latter behaves like $t^{-3}$. 
In our calculation, we have chosen a set of parameters $k_0 =0.0$ and $x_0=0.0$ 
for the three initial states, 
and $a/a_0 =-2.0$ and $b/a_0 =2.0$ for the interval $[a, b]$ for $P(t)$. 
Then, as is seen from Figs.\ \ref{fig:figure1} and \ref{fig:figure2}, 
$P(0) \sim 1$, i.e., the initial states are well localized in the interval.

In conclusion, 
we have considered for every position the long time behavior of the wave packet 
moving freely in one dimension, 
according to the characteristics of the initial wave packet at small momentum.  
We then have found that 
the asymptotic power of $t$ obeyed by 
the wave packet is constant everywhere, 
at most excluding one position $\xi_0$. 
We also have obtained the asymptotic behavior of the nonescape probability at long times, 
and compared that of the survival and nonescape probabilities. 
It is of interest that they can decrease in the different power laws 
depending on the initial states, 
in spite of the apparent similarity between their physical meanings. 
Our derivation can be easily extended to an arbitrary dimension, 
by starting with Eq.\ (\ref{eqn:30}) in a corresponding dimension. 
In these analyses, we assume that the exchange of the order of summation 
and integration is admitted in the formal expansions in 
Eqs.\ (\ref{eqn:70}) and (\ref{eqn:200}). 
Indeed, this assumption can be rigorously guaranteed, 
when we make the same discussion with the finite series involving 
an appropriate remainder, 
instead of Eqs.\ (\ref{eqn:70}) and (\ref{eqn:200}). 
In any such procedure to keep the validity of the formula, e.g., (\ref{eqn:125}), 
what should be satisfied at least is that 
all of the differential coefficient $\hat{\psi}^{(j)}(0)$ with $j$ up to $m$ 
(or $m+1$) is finite for even $m$ (or odd $m$). 
See Eq.\ (\ref{eqn:340}) [or (\ref{eqn:330})]. 
It should be noted that this condition also implicitly implies that 
$\lim_{k\to +0}\hat{\psi}^{(j)}(k) 
=\lim_{k\to -0}\hat{\psi}^{(j)}(k)$ for $j=0, 1, \ldots, m$ (or $m+1$). 
These conditions are satisfied by those $\psi$'s who 
are rapidly decreasing functions as in Eq.\ (\ref{eqn:130}).  
However, such a circumstance is not always valid for an arbitrary initial wave packet, 
e.g., the wave packet with the power-law tail \cite{Unnikrishnan,Lillo,Mendes} or 
that treated in Ref.\  \cite{Damborenea}. 
The former causes $|\hat{\psi}(0)| =\infty$ and the latter does 
$\lim_{k\to +0}\hat{\psi}^{(m)}(k) \neq \lim_{k\to -0}\hat{\psi}^{(m)}(k)$. 
It is then significant to consider how our results are modified for 
such initial wave packets. 
Furthermore, it is important to extend our consideration 
to the potential systems. 
In particular, it is relevant to examine in that case the possible influence of 
the characteristics of the initial states in the long time behavior of 
the survival and nonescape probabilities. 
In fact, such an attempt has not been done in previous investigations. 
An extension may be realized by starting, instead of Eq.\ (\ref{eqn:70}), 
with the asymptotic expansion 
of the wave packet at long times for the short-range potential systems, 
attained by several methods 
(see, for example, Refs.\ \cite{Garcia,Muga,Dijk,Rauch} and references therein).



The author would like to thank Professor I.\ Ohba 
and Professor H.\ Nakazato for useful and helpful discussions, 
and Professor J.\ G.\ Muga and the referees for valuable comments.






\begin{thebibliography}{99}
\bibitem{SP} One may feel that the terms, ``survival'' and ``nonescape'',  
are not appropriate for a free system. However, as is seen in 
Eqs.\ (\ref{eqn:3}) and (\ref{eqn:6}), the definitions of $S(t)$ and $P(t)$ 
are mathematically natural extensions of them for the potential systems, 
and their physical meanings are obvious.
\bibitem{Gj} 
Actually, $G_j$ is the expansion coefficient of the free resolvent 
$
(H_0 -z)^{-1}$ at small $z$ 
\cite{Miyamoto}.
\end{thebibliography}
\end{document}